\documentclass[usenatbib]{basi}
\usepackage[T1]{fontenc}
\usepackage[british]{babel}
\usepackage[varg]{txfonts}
\usepackage{rotating}
\usepackage{dcolumn}
\def\aj{AJ}

\def\apj{ApJ}

\def\apjs{ApJS}

\def\aap{A\&A}

\def\mnras{MNRAS}

\def\sun{\odot}

\begin{document}
\title[Seyferts on Kpc and Pc scales]{Probing Radio Emission in Seyfert Galaxies on Parsec- and Kiloparsec-scales}
\author[P. Kharb et al.]{P. Kharb$^1$\thanks{email: \texttt{kharb@iiap.res.in}}, V. Singh$^2$, J.~F. Gallimore$^3$, C.~H. Ishwara-Chandra$^4$\\
       $^1$Indian Institute of Astrophysics, Koramangala, Bangalore 560034, India\\
       $^2$Institut d'Astrophysique Spatiale, Batiment 121, University Paris Sud XI, 91405 Orsay Cedex, France \\
       $^3$Bucknell University, Lewisburg, PA 17837, USA\\
       $^4$National Centre for Radio Astrophysics - TIFR, Ganeshkhind P.O., Pune 411007, India}
\pubyear{2014}
\volume{00}
\pagerange{\pageref{firstpage}--\pageref{lastpage}}
\date{Received --- ; accepted ---}
\maketitle
\label{firstpage}

\vspace{-0.6cm}
\begin{abstract}
Seyfert galaxies have traditionally been classified as radio-quiet active galactic nuclei. A proper consideration of the nuclear optical emission however proves that a majority of Seyferts are radio-loud. Kpc-scale radio lobes/bubbles are in fact revealed in sensitive observations at low radio frequencies of several Seyferts. Through the use of very long baseline interferometry, we have been able to determine the direction of the parsec-scale jets in some of these Seyfert galaxies. The misalignment between the parsec-scale jets and the kpc-scale lobes that is typically observed, is either suggestive of no connection between the two, or the presence of curved jets that power the radio lobes. In this context, we briefly discuss our new low radio frequency GMRT observations of two Seyfert galaxies with lobes.
\end{abstract}

\vspace{-0.5cm}

\section{Introduction}
\vspace{-0.3cm}
Active galactic nuclei (AGN) are the centres of a small fraction of galaxies that harbour actively accreting supermassive black holes (SMBHs; $10^7-10^9~M_\sun$). Only about $15\%-20\%$ of these have relativistic outflows that are ejected from the BH-accretion disk systems which propagate to 10s$-$100s of kiloparsecs, way beyond the extents of their host galaxies \citep{Kellermann89}. These AGN are referred to as radio-loud. The vast majority of AGN on the other hand, lack kpc-scale collimated radio outflows, making them radio-quiet. The formal quantitative definition for radio-loudness provided by \citet{Kellermann89} is that the ratio ($R$) of the radio flux density at 5 GHz must exceed the optical $B$-band flux density at 4400$\AA$~by~$\ge10$. Seyfert galaxies have historically been identified as radio-quiet AGN (i.e., they have $R<10$). Since radio-loud AGN are typically hosted by large elliptical galaxies, and Seyferts mostly reside in spiral or lenticular  galaxies, explanations for the kpc-scale radio emission have focused on differences in the host galaxy types: for instance, differences in black hole masses and spins, both of which are different in spiral and elliptical galaxies, have been suggested to explain the differences in radio properties \citep[e.g.,][]{Sikora07}. The inherent assumption in the estimation of $R$ however, is that the $B$-band luminosity is dominated by the AGN. This is unlikely to be true for Seyfert galaxies. \citet{HoPeng01} extracted the optical nuclear luminosities for a large number of Seyfert galaxies through a proper modelling of their galactic bulge emission in high resolution {\it HST} images. Remarkably, after taking into account their optical nuclear luminosities, Ho \& Peng found that a majority of the Seyfert galaxies were radio-loud. \citet{Kharb14} estimate that $\ge70\%$ of Seyfert galaxies belonging to the Extended 12$\mu$m Seyfert sample \citep{Rush93} are radio-loud once their optical nuclear emission is considered.
Furthermore, sensitive radio observations at low radio frequencies have revealed the presence of radio structures of typical extents 1$-$10 kpc in several Seyfert galaxies. \citet{Gallimore06} found that $>44\%$ of Seyferts belonging to the {\it complete} CfA+12$\mu$m sample exhibited kpc-scale radio structures when observed with the Very Large Array in the (sensitive) D-array configuration. These structures typically appear like edge-brightened lobes. However, they differ from the lobes of powerful Fanaroff-Riley II (FRII) sources in that they usually do not show collimated jets leading into them or hotspots at their leading edges. Seyfert lobes raise interesting questions. Are they powered by AGN jets? If yes, do very long baseline interferometry (VLBI) observations reveal parsec-scale radio jets in Seyferts, similar to what is observed in radio-loud AGN ? Are these parsec-scale jets aligned (or not) with the kpc-scale radio lobes? These are some of the questions that we have been attempting to address in our study of Seyfert galaxies with lobes.

\section{Observations on Kiloparsec and Parsec-scales}
\vspace{-0.3cm}
{\bf A look at a few individual Seyferts:}
Phase-referenced VLBI observations at 1.6 and 4.9~GHz of two Seyfert galaxies with lobes, namely NGC6764 and Mrk6, have revealed the presence of parsec-scale core-jet structures in them \citep{Kharb10a,Kharb14}. These jets show a large misalignment ($50^\circ-60^\circ$) with the kpc-scale lobes. Parsing through the literature and the NRAO image archive\footnote{https://archive.nrao.edu/archive/archiveimage.html} we find that such large jet misalignments are also observed in NGC3079, NGC7212, Ark564, among others \citep[see VLBI images in][]{Sawada00,Lal04}. This finding could either suggest that the parsec-scale jets are not associated with the kpc-scale lobes, or they are curved and do indeed connect to and power the radio lobes. However as we see below, the former scenario is disfavoured by observations. Mrk6 possesses not just one set of radio lobes but two of them aligned perpendicular to each other \citep{Kharb06}. A flip in the jet direction powering the lobes can explain both sets of lobes. Star-formation rates (SFR) derived from {\it Herschel} (L250$\mu$m) data imply a SFR $<0.8~M_\sun$~yr$^{-1}$ in Mrk\,6. This is much lower than the SFR of $\sim33~M_\sun$~yr$^{-1}$ expected if the lobes were inflated by starburst superwinds. {\it Chandra} observations of Mrk6 and NGC6764 indicate that the X-ray emitting gas in these Seyferts is shock-heated by an AGN outflow \citep{Mingo11,Croston08}. Therefore, in these Seyfert galaxies and others, there are indications that the radio lobes are AGN outflow rather than starburst wind driven. Curved radio jets that could be powering the Seyfert lobes may arise due to precession of the central engine or the galaxy ISM rotation along with strong jet-medium interaction. 

\vspace{-0.13cm}
\noindent {\bf GMRT observations of Seyferts:} In order to image the typically steep-spectrum radio lobes, we have observed two Seyfert galaxies, {\it viz.,} NGC4235 and NGC4594 (the Sombrero), with the GMRT at 325 and 610 MHz. Radio lobes that were previously indicated in the 5~GHz VLA observations of \citet{Gallimore06} are also detected with the GMRT. A preliminary radio image of NGC4594 at 610 MHz is presented in Fig.1 (Kharb et al., 2014, in prep.). Apart from the lobes (top left panel, Fig.1), radio emission coincident with the host galaxy is clearly seen in the east-west direction (top right and bottom panels, Fig.1). We are currently in the process of analysing the 325~MHz data and creating the spectral index maps for these Seyferts. The 8.4~GHz VLBI images of these galaxies that are available in the literature reveal unresolved cores in both \citep{Anderson04,Ojha05}. However, since lower frequency VLBI data exists in the NRAO archive, we plan to obtain greater information on the parsec-scale jets in these sources. 

\section{Summary}
\vspace{-0.3cm}
A proper consideration of the nuclear optical emission in Seyfert galaxies reveals that the majority of them are radio-loud. Sensitive radio observations of Seyferts reveal the presence of kpc-scale lobes in them. VLBI observations detect parsec-scale jets in Seyferts that are misaligned by large degrees ($50^\circ-60^\circ$) to the kpc-scale lobes. This could either indicate that the jets are not associated with the lobes or that they are curved and indeed power the lobes. The former scenario is disfavoured by several observational findings that indicate that the lobes are AGN outflow rather than starburst wind driven. Curved radio jets could arise due to precession of the central engine or host galaxy ISM rotation and jet-medium interaction. We are currently in the process of analysing 325 and 610~MHz data from the GMRT of two Seyfert galaxies with lobes, {\it viz.,}, NGC4235 and NGC4594. The spectral ageing analysis that we plan to carry out in the near future will reveal the age of the lobes in these Seyferts, while an analysis of archival VLBI data will reveal the structure of their parsec-scale jets.

\begin{figure}
\includegraphics[trim = 0mm 12mm 0mm 0mm, width=5.5cm]{fig2.ps}
\hspace*{-0.07cm}\includegraphics[trim = 5mm 12mm 0mm 0mm, width=10.3cm]{fig3.ps}
\hspace*{2.2cm} \includegraphics[trim = 0mm 12mm 0mm 0mm, width=10.3cm]{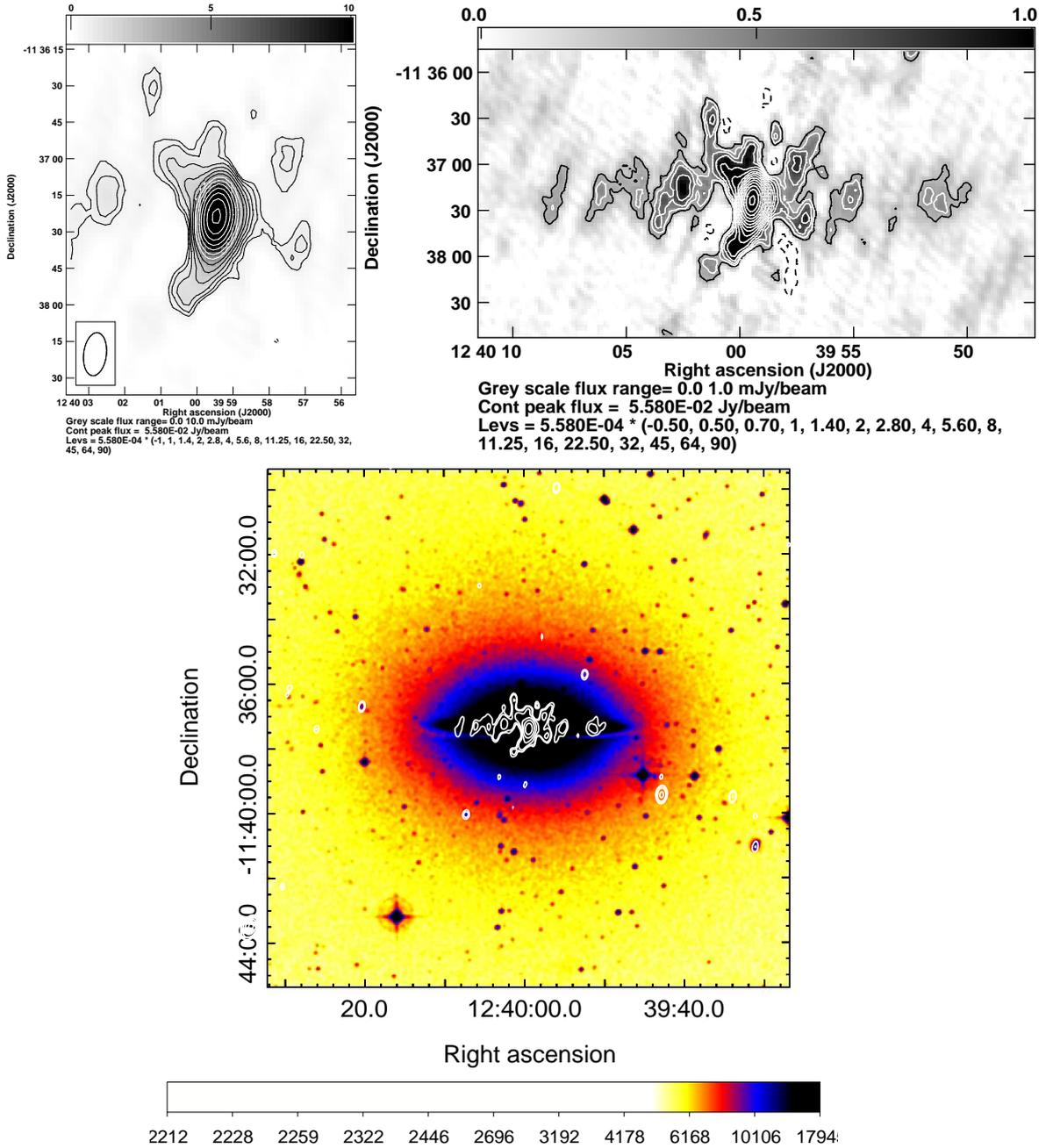}
\caption{Top panels: Preliminary 610~MHz contour images of NGC4594 (the Sombrero galaxy) observed with the GMRT. Different contour levels are chosen to highlight the Seyfert lobes (top left) and the lobe+galactic radio emission (top right). Bottom panel: 610~MHz radio contours in white superimposed on the DSS optical image of the host galaxy (Kharb et al., 2014, in prep.).}
\end{figure}

\label{lastpage}


\vspace{-0.3cm}
\begin{thebibliography}{}
{\small
\bibitem[\protect\citeauthoryear{{Anderson} et~al.}{{Anderson} et~al.}{2004}]{Anderson04}
{Anderson} J. M.,  {Ulvestad} J.~S., {Ho} L. C., 2004, \apj, 603, 42
\bibitem[\protect\citeauthoryear{{Croston} et~al.}{{Croston} et~al.}{2008}]{Croston08}
{Croston} J. H., {Hardcastle} M. J., {Kharb} P., {Kraft} R., {Hota} A., 2008, \apj, 688, 190
\bibitem[\protect\citeauthoryear{{Gallimore} et~al.}{{Gallimore} et~al.}{2006}]{Gallimore06}
{Gallimore} J. F., {Axon} D.~J.,  {O'Dea} C. P.,  {Baum} S. A., {Pedlar}
  A.,  2006, \aj, 132, 546
\bibitem[\protect\citeauthoryear{{Ho} \& {Peng}}{{Ho} \&
  {Peng}}{2001}]{HoPeng01}
{Ho} L.~C.,  {Peng} C.~Y.,  2001, \apj, 555, 650
\bibitem[\protect\citeauthoryear{{Kellermann} et~al.}{{Kellermann} et~al.}{1989}]{Kellermann89}
{Kellermann} K.~I.,  {Sramek} R.,  {Schmidt} M.,  {Shaffer} D.~B.,    {Green}
  R.,  1989, \aj, 98, 1195
\bibitem[\protect\citeauthoryear{{Kharb} et~al.}{{Kharb} et~al.}{2006}]{Kharb06}
{Kharb} P.,  {O'Dea} C. P.,  {Baum} S. A.,  {Colbert} E.,  {Xu} C.,
  2006, \apj, 652, 177
\bibitem[\protect\citeauthoryear{{Kharb} et~al.}{{Kharb} et~al.}{2010}]{Kharb10a}
{Kharb} P.,  {Hota} A.,  {Croston} J. H.,  {Hardcastle} M. J.,  {O'Dea} C. P., et~al., 2010, \apj, 723, 580
\bibitem[\protect\citeauthoryear{{Kharb} et~al.}{{Kharb} et~al.}{2014}]{Kharb14}
{Kharb} P.,  {O'Dea} C. P.,  {Baum} S. A.,  {Hardcastle} M. J., et al.,
  2014, \mnras, accepted, arXiv:1402.7174
\bibitem[{{Lal} {et~al.}(2004){Lal} et~al.}]{Lal04}
{Lal} D., {Shastri} P., {Gabuzda} D., 2004, \aap, 425, 99
\bibitem[\protect\citeauthoryear{{Mingo} et~al.}{{Mingo} et~al.}{2011}]{Mingo11}
{Mingo} B.,  {Hardcastle} M. J.,  {Croston} J. H.,  {Evans} D.,  {Hota} A., {Kharb} P., et al., 2011, \apj, 731, 21
\bibitem[\protect\citeauthoryear{{Rush} et~al.}{{Rush} et~al.}{1993}]{Rush93}{Rush} B., {Malkan} M.~A.,    {Spinoglio} L.,  1993, \apjs, 89, 1 
\bibitem[\protect\citeauthoryear{{Ojha} et~al.}{{Ojha} et~al.}{2005}]{Ojha05}
{Ojha} R., {Fey} A., {Charlot} P., {Jauncey} D., {Johnston} K., 2005, \aj, 130, 2529
\bibitem[\protect\citeauthoryear{{Sawada-Satoh} et~al.}{{Sawada-Satoh}
  et~al.}{2000}]{Sawada00}
{Sawada-Satoh} S.,  {Inoue} M.,    {Shibata} K., et~al., 2000, PASJ, 52, 421 
\bibitem[\protect\citeauthoryear{{Sikora} et~al.}{{Sikora}
  et~al.}{2007}]{Sikora07}
{Sikora} M.,  {Stawarz} {\L}.,  {Lasota} J.-P.,  2007, \apj, 658, 815   }
\end{thebibliography}
\end{document}